\documentclass[aps,showpacs,toolkits]{revtex4}
\usepackage{epsf}
\usepackage{amssymb}
\usepackage{graphicx}

\begin{document}

\title{\Large \bf   Holographic  Wave Functions, Meromorphization  
and Counting Rules} 

\author{A.V. RADYUSHKIN\footnote{Also  at Bogoliubov Laboratory of Theoretical Physics, JINR, Dubna, Russian Federation}}

\affiliation{ Physics Department, Old Dominion University, 
 Norfolk, VA 23529, USA \\ 
 and  \\
 Theory Center, Jefferson Lab,  Newport News, VA 23606, USA}

\begin{abstract}
We 
 study the large-$Q^2$ behavior of the 
 meson form factor $F_M (Q^2)$   constructed 
 using the holographic light-front wave functions 
 proposed recently by Brodsky and de Teramond.
 We show that this  model can be  also  obtained  within the Migdal's 
 regularization 
  approach (``meromorphization''), 
  if one  applies it to  3-point function for scalar currents made of scalar quarks. 
 We found that the asymptotic $1/Q^2$ behavior 
of  $F_M (Q^2)$   is generated by soft  Feynman mechanism rather than by 
large transverse momentum dynamics, which 
 causes  very late onset of the 
  $1/Q^2$ asymptotic behavior. It becomes visible 
  only for   unaccessible momenta $Q^2 \gtrsim 10$\, GeV$^2$.
  Using meromorphization for  spin-1/2 quarks, we demonstrated that 
 resulting form factor $F^{\rm spinor}_M (Q^2)$ 
 has $1/Q^4$ asymptotic 
 behavior. Now, owing to the  late onset of this 
  asymptotic  pattern,  $F^{\rm spinor}_M (Q^2)$
 imitates the $1/Q^2$ 
  behavior in the few GeV$^2$ region.  
  We discuss  analogy between  meromorphization and local quark-hadron 
 duality model for the  pion form factor, and show that    
 adding  the ${\cal O}(\alpha_s)$ correction 
  to the spectral function  
  brings in  the hard pQCD  contribution that 
  has the dimensional counting $1/Q^2$ behavior
  at large $Q^2$. At  accessible $Q^2$, 
  the   ${\cal O}(\alpha_s)$ term  is a rather 
  small fraction of the total result. 
  In this scenario,  the  ``observed'' quark counting 
  rules  for hadronic form  factors  is an approximate 
  phenomenon resulting   from 
  Feynman mechanism in its preasymptotic regime.  
  
  \end{abstract} 
  \pacs{11.15.Tk, 11.55.Hx, 12.38.-t, 12.39.St, 12.40Yx, 13.40.Gp, 14.40.-n}

 \maketitle

\section{Introduction} 

Experimental evidence  that form factors of hadrons
consisting of $n$ quarks have behavior close to $1/(Q^2)^{n-1}$,
provokes  expectations  that there is a fundamental and/or easily visible 
reason for such a phenomenon, scale invariance being   the 
 most natural suspect. 
 Indeed, for an elementary fermion, 
 EM form factor is  constant, 
 $\langle P' |J^{\mu} |P \rangle = \bar u(P') \gamma^{\mu} u(P) $.
 When $(n-1)$ spectator quark fields with dimension 
 (mass)$^{3/2}$ are added to the initial  state   $|P \rangle$  and the final state 
 $\langle P' | $,  
the extra  kinematical  factors  $\sim [u(P) \bar u(P')]^{n-1}$ 
  take  only  worth of  (mass)$^{(n-1)}$, 
and  a constant of dimension (mass)$^{2(n-1)}$
is needed to  take care of the rest. If, except for this overall constant,
 no other dimensionful 
constants can show up   for large $Q^2$,  then  
form factor has the quark counting rule  $1/(Q^2)^{n-1}$ behavior 
\cite{Matveev:1977ra}. 
A specific dynamical mechanism \cite{Brodsky:1973kr} that produces a   scale 
invariant behavior is provided by hard rescattering in a theory with 
dimensionless coupling constant.  
After advent of QCD, pion and nucleon form factors
were calculated within the hard scenario
\cite{Farrar:1979aw,Chernyak:1977fk,Radyushkin:1977gp,Lepage:1979zb},
and it was realized  that  
perturbative QCD predicts, in fact, the $(\alpha_s/Q^2)^{n-1}$
asymptotic behavior.  
In  the pion case, the prediction is 
$F_\pi (Q^2) = (2\alpha_s/\pi)s_0/Q^2$, where 
$s_0 = 4 \pi^2 f_\pi^2 \approx 0.7$\,GeV$^2$ 
is a constant close to  $m_\rho^2\approx 0.6$\,GeV$^2$. 
This indicates  that the pQCD asymptotics 
{\it is not} the large-$Q^2$ limit of the phenomenologically successfull 
VMD fit   $F_\pi^{\rm VMD} (Q^2) \sim 1/(1+Q^2/m_\rho^2)$,
but rather looks like  ${\cal O} (\alpha_s)$ correction to it.
Also, the  smallness of $\alpha_s/\pi$ 
undermines  attempts to describe    
available data solely by pQCD hard mechanism. 
During the last years, the growing consensus 
is that at available $Q^2$, form factors are dominated 
by soft contributions described by   nonforward parton 
densities ${\cal F} (x, Q^2)$ \cite{Radyushkin:1998rt} (or
 generalized parton distributions $H(x,\xi;Q^2)$ for zero skewness $\xi$),
and successful fits were  obtained 
\cite{Belitsky:2003nz,Diehl:2004cx,Guidal:2004nd}
in models with 
 ${\cal F} (x, Q^2) = f(x) \, e^{-Q^2 g(x)}$ 
 having exponential behavior  for large $Q^2$
at fixed $x$. If   $g(x)$ vanishes
for $x \to 1$ like $ (1-x)^a$, a powerlike asymptotics 
$F(Q^2) \sim (1/Q^2)^{(b+1)/a}$ 
in this case appears only after integration over 
$x$, i.e., it is governed by the   
 so-called  Feynman mechanism \cite{Feynman}, 
 and is determined by the $x \to 1$ behavior $f(x) \to  (1-x)^b$
 of the  preexponential function $f(x)$, in contrast to the 
 hard mechanism for which the subprocess amplitude already  has the 
 $(1/Q^2)^{n-1}$ power behavior not affected by  subsequent
 $x$-integrations.

Another new development 
is related to applications of    AdS/CFT construction 
to QCD  and claims \cite{Polchinski:2002jw,Brodsky:2003px}
 that this framework (often  referred as ``AdS/QCD'')
provides a nonperturbative explanation 
of  quark counting rules.    
 In this 
scenario, they  reflect the conformal invariance of the 5-dimensional theory,
in particular, the power behavior of the normalizable modes
$\Phi  (\zeta)$ at small values of the 5th coordinate 
$\zeta$. Parametrically, the prediction has the form 
$(\Lambda^2/Q^2)^{n-1}$, 
without any accompanying $\alpha_s^{n-1}$ factors.
Given  explicit expressions for  $\Phi  (\zeta)$, 
with the  value of  $\Lambda$
fixed from fitting the hadron masses,
it is straightforward to check the structure 
of AdS/QCD results  for form factors
and their potential  to describe the features of existing data.
This is one of the goals of  the present investigation. 
Another is to study the recently proposed 
interpretation \cite{Brodsky:2006uq} of  AdS/QCD results 
in terms of light-front wave functions,
which opens a possibility to find out whether, 
in terms of the light-cone momenta $x, {\bf k}_{\perp}$,  
the AdS/QCD quark counting corresponds 
to large-${\bf k}_{\perp}$ hard 
mechanism or, as we will show,
 to  the $x \to 1$ soft Feynman/Drell-Yan \cite{Feynman,Drell:1969km} mechanism. 
In view of yet another recent observation \cite{Erlich:2006hq} that
some of the results of the AdS/QCD 
approach coincide with  those of Migdal's program
\cite{Migdal:1977nu} (that starts with a perturbative 
correlator and substitutes its cuts by hadron 
poles),  we apply the  extension \cite{Dosch:1977qh}
of this  ``meromorphization'' idea 
 to the 3-current correlators, and establish a
 connection between this approach and holographic light-front
 wave functions of Ref. \cite{Brodsky:2006uq}.

Finally,  we  discuss  analogy  between the meromorphization 
procedure and the ``local quark-hadron duality'' model
\cite{Nesterenko:1982gc}
that succesfully describes the pion form factor data
and gives a unified description of its  soft 
and hard parts. The soft term $F_\pi^{\rm LD} (Q^2)$ 
in this approach dominates at accessible
$Q^2$, but 
has the $1/Q^4$ asymptotic behavior. Due to its late  onset,
the curve $Q^2 F_\pi^{\rm LD} (Q^2)$  
has a wide plateau in a few GeV$^2$ region, 
i.e., $F_\pi^{\rm LD} (Q^2)$ 
imitates there the   $1/Q^2$  behavior.
Thus, the desired  $1/Q^2$  result for accessible $Q^2$ 
is obtained because  the nonperturbative term 
has a faster, $1/Q^4$  asymptotic fall-off. 
On the other hand, the curve  $Q^2 F_M (Q^2)$ 
for the meson form factor in the holographic model
of Ref. \cite{Brodsky:2006uq} monotonically increases to its asymptotic value
$\sim 2.6 \, m_\rho^2$,
and is far from being flat in the whole accessible region 
$Q^2 \lesssim 10$\,GeV$^2$.

 \section{ Holographic wave functions}   
 
 We will need  some 
 elements of the derivation of the holographic wave functions 
 proposed in Ref. \cite{Brodsky:2006uq}. In the hard-wall approximation
 \cite{Polchinski:2002jw}, 
the expression  
for the elastic form factor in the holographic duality model is given by 
 \begin{equation} 
F(Q^2)
= \int_0^{1/\Lambda} \frac{d\zeta}{\zeta^{3}} \, 
 \Phi_{P'}(\zeta) J(Q,\zeta) \Phi_P(\zeta) \ , 
\label{AdS}
\end{equation}
 where   $J(Q,\zeta) = \zeta Q K_1 (\zeta Q)$ is the ``nonnormalizable'' mode
 describing the EM current, and  $\Phi_P(\zeta)$, 
 $\Phi_{P'}(\zeta)$ come from  the ``normalizable'' modes
 describing    initial and final states, with 
 $\Phi(\zeta) = C \zeta^2 J_L(\beta_{L,k} \zeta \Lambda)$  
 for a $\bar q q$ state with orbital momentum $L=0,1, \ldots $ 
 and radial number $k=1,2, \ldots$, and  
 $ \beta_{L,k}$ being  the $k$th root of the Bessel function 
 $J_L(x)$. 
 On the other hand, in the light-cone (LC) formalism,   
\begin{eqnarray} 
F(Q^2)=\int^1_0 dx  \int d^2 {\bf \eta}_\perp 
e^{i  {\bf \eta}_\perp \cdot {\bf q}_\perp}
{\cal B} ( x, {\bf \eta}_\perp  )   
 \equiv \int^1_0 dx \, {\cal F} (x,Q^2) ,
\label{Bxeta}
\end{eqnarray}
where   ${\cal B} ( x, {\bf \eta}_\perp  )$ is a parton density 
function  \cite{Soper:1976jc}
that depends  on $x$,  the 
light-cone momentum fraction  of  the active quark, 
and 
 ${\bf \eta}_\perp$,   the 
  $x_i$-weighted transverse position of spectators.  
  The function  ${\cal B} ( x, {\bf \eta}_\perp  )$ 
accumulates information from all 
Fock components 
and its $ {\bf \eta}_\perp$   Fourier transform  gives the generalized parton 
distribution (GPD) ${\cal F} (x,Q^2)$.  In particular,
 the   2-body  part  
 of  a meson form factor is given by 
\begin{equation} 
F_{(2)}(Q^2) =\int^1_0 dx  \int d^2 {\bf b}_\perp 
e^{i \bar x {\bf b}_\perp \cdot {\bf q}_\perp}
\Bigl |\Psi_2 (x,{\bf b}_\perp) \Bigr |^2 \  ,
\label{bperp}
\end{equation}
where $\bar x \equiv 1-x$. 
With the wave function depending  
on  ${\bf b}_\perp$    through $b\equiv  |{\bf b}_\perp|$ one obtains 
\begin{eqnarray} 
F_{(2)} (Q^2) = 2 \pi \int^1_0 dx  \int_0^{\infty}  b d  b
J_0 (\bar x  b Q) 
|\Psi_2 (x,b) |^2\label{FJ0}  =  2 \pi \int^1_0 
 \frac{dx}{x \bar x} \int_0^{\infty}  z  d  z 
J_0 \left (\sqrt{\frac{\bar x}{x}} \,  z \, Q \right ) \, 
|\phi (x,z) |^2 \  , 
\end{eqnarray} 
where       $z \equiv \sqrt{x \bar x} b$ 
and  wave function was written  as $\Psi_2 (x,b) = \phi (x,z)$. 
Noticing that the integral 
\begin{equation} 
\int_0^1 dx J_0\left( \sqrt{\frac{1-x}{x}} \,  z \, Q \right) = 
z Q K_1(z Q) \equiv {\cal K}_1 (zQ)  = J(Q,z)
\label{J0int}
\end{equation}
gives  $  J(Q,z)$, and assuming that  
 $|\phi (x,z) |^2 = x\bar x \chi^2(z)$,  
 LC  formula (\ref{FJ0}) is cast   into   the form 
\begin{equation} 
F(Q^2) =2 \pi   \int_0^{\infty}  z  d  z \  
 {\cal K}_1 (zQ) \, 
\chi^2(z) 
\label{LCaDS}
\end{equation}
that converts into  
Eq. (\ref{AdS}) if one identifies 
$\Phi (\zeta) \sim \zeta^2 \chi (\zeta)$.   
In   general case,   $z$ is
introduced   through  
 $|\eta_{\perp}| = z \sqrt{\bar x /x}$, and   one assumes  that
 ${\cal B} ( x, {\bf \eta}_\perp  )  = (x/ \bar x) \chi^2 (z)$. 
 This gives the  ``holographic'' model \cite{Brodsky:2006uq} 
  for  light-front wave functions. By construction,
  they should guarantee the $F_M (Q^2=0)=1 $ constraint, i.e.,
  they are  
 {\it effective} wave functions (see, e.g., \cite{Radyushkin:1995pj}) 
normalized by 
\begin{equation} 
\int^1_0 dx  \int d^2 {\bf b}_\perp 
|\Psi_{\rm eff}  (x,{\bf b}_\perp) |^2 \  = 1   \ ,  
\label{Prob}
\end{equation}   unlike   the two-body  wave functions 
$\Psi_2 (x, {\bf b}_{\perp })$ 
normalized 
to   the probability  of  finding the  meson in the   
$\bar q q$  Fock state.  
We will focus on  the lowest meson having  $L=0,k=1$
and mass $M= \beta_{0,1} \Lambda$.  Then 
\begin{equation} 
\Psi_M (x,b) = 
\frac{M \sqrt{ x \bar x/\pi}}
{\beta_{0,1} J_1(\beta_{0,1})} \, J_0(
\sqrt{x \bar x} M b ) \, \theta (\sqrt{x \bar x} b \leq \beta_{0,1}/M)
\ ,  
\label{Psi0}
\end{equation}
where $\beta_{0,1}J_1(\beta_{0,1} ) \approx 1.2$.
 The  ${\bf k}_\perp$ counterpart of this 
 wave function is given by 
 \begin{equation} 
\widetilde \Psi_M (x, {\bf k}_\perp)  =  \frac1{2\pi} 
\int d^2 {\bf b}_\perp 
e^{-i {\bf b}_\perp \cdot {\bf k}_\perp}
\Psi_M (x,{\bf b}_\perp) = \frac{ M}{\sqrt{\pi x \bar x} }  \,
\frac{J_0 (\beta_{0,1} k_\perp/\sqrt{x \bar x} M)}
{M^2 - {k_\perp^2}/{x \bar x }}  
 \  .
\label{kperp}
\end{equation}

\begin{figure}[ht]
 \mbox{\epsfysize=4cm
  \hspace{0cm}
  \epsffile{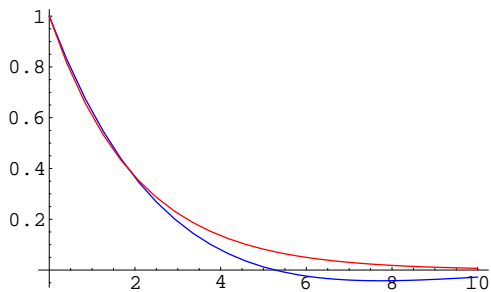}}
 \caption{ Momentum wave function $M \sqrt{\pi x \bar x } \,  \widetilde \Psi_M (x, {\bf k}_\perp)$
(lower curve, blue online) as a function of ${k_\perp^2}/M^2 {x \bar x }$;
 for comparison also shown 
$\exp ( - {k_\perp^2}/2M^2 {x \bar x })$ (upper curve, red online). 
 }
\label{momWF}
\end{figure}

Since $\beta_{0,1}$ is a root of  $J_0(z)$,
 the momentum wave function $\widetilde \Psi_M (x, {\bf k}_\perp)$ 
 has no singularities for $ {k_\perp^2}/{x \bar x } = M^2$,
 and  it has zeros when   $ {k_\perp^2}/{x \bar x }$ coincides 
 with squared masses of higher states. 
For large $k_\perp$, wave function  oscillates with the magnitude 
decreasing as $1/k_\perp^{5/2}$,


 \begin{equation} 
 \widetilde \Psi_M (x, {\bf k}_\perp) \sim  \,  - \frac{M}{{\pi k_\perp^2}} \, 
 \sqrt{2x \bar x} \ 
 \frac{\cos (k_\perp/\sqrt{x \bar x} \Lambda-\pi/4)}
 {\sqrt{k_\perp/\sqrt{x \bar x} \Lambda}}
 \  .
\label{psikas}
\end{equation}
    This result  contradicts  the statement made   
    in  Ref. \cite{Brodsky:2003px} 
that the AdS/QCD construction corresponds to a purely 
power-law large-$k_\perp$ behavior $\psi (k_\perp) \sim (1/k_\perp^2)^{n-1}$ 
for the wave function of $n$-particle bound state.

\section{Feynman mechanism }

\begin{figure}[h]
 \mbox{\epsfysize=4cm
  \hspace{0cm}
  \epsffile{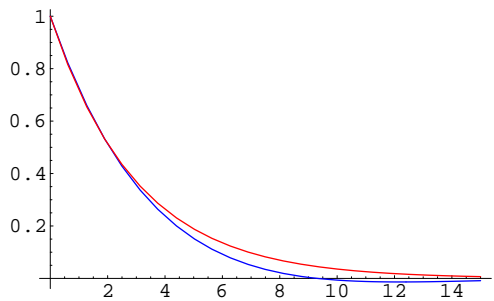}}  
\caption{ Generalized parton distribution 
${\cal F}_M(x,Q^2)$ (lower curve, blue online)
 as a function of $\bar x Q^2/xM^2 \equiv  \sigma$.
 For comparison, we  also show 
$\exp (- \sigma/3)$  (upper curve, red online). 
 }
\label{GPD}
\end{figure}

We can  get the  large-$Q^2$  asymptotics 
of  the lowest  state form factor 
 \begin{equation} 
F_M(Q^2)
=
 \frac{2M^2}{Q^2  [\beta_{0,1}J_1(\beta_{0,1})]^2}
  \int_0^{Q/\Lambda} \xi \,  {d\xi} \, 
{\cal  K}_1 (\xi ) \, J^2_0(\xi M /Q ) 
 \  ,
\label{F0AdS}
\end{equation}
 by  Taylor expanding    $ J_0^2(\xi M /Q )$ 
 and   
neglecting ${\cal O} ( e^{-Q/\Lambda})$  
terms from the upper limit 
of $\xi$ integration: 
\begin{equation} 
F_M(Q^2)
=  \frac{4M^2}{Q^2  [\beta_{0,1}J_1(\beta_{0,1})]^2}
 \left [ 1-  \frac{4M^2}{Q^2} 
+\frac98  \left (\frac{4 M^2}{Q^2} \right )^2 
+ {\cal O} (M^6/Q^6) \right ]+ {\cal O} (e^{-Q/\Lambda})
 \sim \frac{0.64}{1+Q^2/4M^2}
\ . \label{FM}
\end{equation}
Though this result has a  monopole-like structure,
the scale $4 M^2$  is  evidently   too large. 
With $M=m_\rho$, 
the curve for  $Q^2  F_M (Q^2)$
 is far from being flat in the  accessible  region  $Q^2 \lesssim 10$\, GeV$^2$.
Note also,  that though  the large-$Q^2$  behavior of $F_M(Q^2)$  
 is determined by  the  small-$\zeta$ behavior 
 of $\Phi (\zeta) $, when  $\zeta$ is  interpreted   
as $\sqrt{x \bar x} b$, the value  
 $\zeta =0$  may correspond to $x=1$ rather than 
 to 
 large transverse momenta. 
Let us  see which  mechanism is 
responsible for the $1/Q^2 $ asymptotics in  terms of  
 the $x, {\bf k}_\perp$ Drell-Yan formula \cite{Drell:1969km}
\begin{equation}
F_M(Q^2) =  \int_0^1 dx \int d^2 {\bf k}_\perp 
\, \widetilde \Psi_M^* (x,{\bf k}_\perp+\bar x {\bf q}_\perp )
\widetilde \Psi_M (x,{\bf k}_\perp)
  \ , \label{FFPsi}
\end{equation}
 $Q \equiv |{\bf q}_\perp|$.
With 
$\widetilde \Psi_M (x,{\bf k}_\perp)$
 decreasing  at large ${\bf k}_\perp$,
  there are two possibilities
 \cite{huang} (see also \cite{Mukherjee:2002gb}): \\
$a)$   finite $x$ and  small $|{\bf k}_\perp |$, e.g., the region 
\mbox{$|{\bf k}_\perp | \ll \bar x  |{\bf q}_\perp|$,}  
where  $\widetilde \Psi_M (x,{\bf k}_\perp)$ 
 is maximal. 
This gives  
  \begin{eqnarray}
F_M(Q^2) \approx   \int_0^1 dx \int d^2 {\bf k}_\perp 
\, | \widetilde \Psi_M^* (x, \bar x {\bf q}_\perp ) \, 
\widetilde \Psi_M (x,{\bf k}_\perp)|
 +
\{ \widetilde \Psi \Leftrightarrow \widetilde \Psi^* \}  \sim    
2 \int_0^1 {dx} \, | \widetilde \Psi_M^* (x, \bar x {\bf q}_\perp ) \, 
\varphi(x)  | \  . 
\end{eqnarray}
  In the hard scenario,  when the $x$  integral is   not  dominated by 
 the $x \to 1$ region, the large-$Q^2$ behavior 
of the form factor repeats 
the large-$k_\perp$ behavior of the momentum wave function.
The latter  does not behave as $1/k_\perp^{2}$: it 
oscillates with magnitude  decreasing  as
$1/k_\perp^{2.5}$, see Eq. (\ref{psikas}),  so  
we need to turn to  the second  possibility: \\
$b)$  $x$ is close to 1, 
so that  $|\bar x {\bf q}_\perp|
\sim |{\bf k}_\perp|$, and $ |{\bf k}_\perp|$  is small.
Then   both   $\widetilde \Psi_M (x,{\bf k}_\perp)$
and $\Psi_M^* (x,{\bf k}_\perp+\bar x {\bf q}_\perp )$ are maximal.
In  Ref. \cite{Drell:1969km}, it was argued that the dominant contribution comes 
from $\bar x |{\bf q}_\perp | \lesssim m = {\rm const}$, and 
the large-$Q^2$ behavior of the form factor
in this scenario 
 reflects 
the phase space available for such configurations.
To  make a specific estimate,  
we represent   the form factor 
as the  $x$-integral   (\ref{Bxeta}) 
of  GPD  
\begin{eqnarray}
{\cal F}_M(x,Q^2)= \frac{2}{\beta^2_{0,1} J_1^2 (\beta_{0,1})}  
 \int_0^{\beta_{0,1}} y  d  y \, 
J_0 \left ( \sqrt{\frac{\bar x}{x}}  \, \frac{Q}{M} \, y \right  ) 
J_0^2 (y) \equiv {\cal G} (\sigma) \,  ,
\label{GPD0}
\end{eqnarray}
which in this case  is a  function ${\cal G} (\sigma)$ of 
$\sigma \equiv \bar x Q^2 /x M^2 $   (see Fig. \ref{GPD}). 
The form factor is then given by  
\begin{equation}
F_M(Q^2) = \frac{M^2}{Q^2} \int_0^{\infty}  d \sigma \,
\frac{{\cal G}(\sigma)}{(1+ \sigma M^2 /Q^2)^2}  \  , 
\end{equation}
and the leading $1/Q^2$ term is proportional 
to zeroth $\sigma$-moment of ${\cal G}(\sigma)$,
which does not vanish though  the function   ${\cal G}(\sigma)$   
oscillates for large values of $\sigma$.
In fact, it   is very small  in the region 
above its first zero (at $\sigma \approx 9$) 
 while 
in the region below this zero ${\cal G}(\sigma)$ 
is very close to 
the  exponential $e^{-\sigma/2.6 } $, the  integration of which 
gives the correct  coefficient for the ${\cal O} (M^2/Q^2)$ 
term.    
Thus,   the leading  large-$Q^2$  term is given by integration 
over $\sigma  \lesssim 10$ or,  returning to the $x$-variable, 
by the region  $\bar x \lesssim   10 M^2/Q^2$
(we see again that asymptotic estimates can be used 
only for $Q^2 \gg 4 M^2$).
The   $\bar x \lesssim {\cal O} ( M^2 /Q^2)$ result parametrically differs 
 from  the   estimate given  in Ref. \cite{Drell:1969km}.  
The reason  is that  the transverse momentum 
${\bf k}_\perp$ enters into the wave function 
through  the combination 
$|{\bf k}_\perp|^2 /x \bar x$, and    
the   $x$-size of the dominant  region
is determined by 
$|{\bf k}_\perp+\bar x {\bf q}_\perp|^2 /x \bar x M^2
 \lesssim  {\rm const}$ or  $|\bar x {\bf q}_\perp|^2 /x \bar x 
 \lesssim {\rm const} \, M^2$, which gives $\bar x  
 \lesssim {\rm const} \, M^2/Q^2$.

 Thus, we found that  the large-$Q^2$ asymptotics of the meson form factor
in the model of Ref. \cite{Brodsky:2006uq}   is 
 governed by the {\it soft}  Feynman mechanism, with the   power-law asymptotics   
   determined by
  the $x \to 1 $  behavior     of the prefactor  $f(x)$  accompanying 
 a  decreasing
 function of $\bar x Q^2/x\Lambda^2$. In the present  case, $f(x)=  1$, 
 which  gives $F_M (Q^2) \sim 1/Q^2$. 
 With extra  $\bar x^N$ factor, 
the outcome 
of the $x$-integration would be  $(\Lambda^2 /Q^2)^{N+1}$.
 Clearly, the prefactor $f(x)$  is nothing else but 
 the parton distribution function:
 \begin{eqnarray}
 f(x) = {\cal F}(x,Q^2=0)= 
  \int \, d^2 {\bf b}_\perp \, 
|\Psi (x,{\bf b}_\perp) |^2
 =  \int \, d^2 {\bf k}_\perp \, 
|\Psi (x,{\bf k}_\perp) |^2\,  ,
\label{PD}
\end{eqnarray}
i.e., the model of Ref. \cite{Brodsky:2006uq} gives   a  constant,  
$x$-independent parton distribution  $f(x) =1$.

 If we approximate  $\widetilde \Psi_M (x,{\bf k}_\perp)$ by a Gaussian
 function  $\widetilde \Psi_G (x,{\bf k}_\perp) \sim 
 \exp [ - {k_\perp^2}/2M^2 {x \bar x }]/ \sqrt{x \bar x} $  
(see Fig. \ref{momWF}),   
  the relevant GPD  ${\cal F}_G (x,Q^2)$ can be calculated
 analytically. It is instructive 
 to perform such a calculation in the impact parameter space.
 Then $ \Psi_G (x,b) \sim \sqrt{x \bar x} 
   \exp [ - {b^2} M^2 {x \bar x }/2]$ (in fact, a Gaussian 
  wave function for a ground state is  obtained 
  in the hologaphic model of Ref. \cite{Karch:2006pv}
  that gives a linear law for (mass)$^2$ of excited states), 
  and integrating  over {\it all} 
 (including small)   values of 
   ${\bf b}_\perp$ in the form factor formula  (\ref{bperp}) 
   gives
   ${\cal F}_G (x,Q^2) = e^{-\bar x Q^2/4xM^2}$,
   a function that, for any fixed $x$,  
   vanishes exponentially at large $Q^2$.
    The power-law asymptotics is obtained only after 
    integrating over $x$, which gives 
    \begin{equation} 
F_G(Q^2) = \sum_{n=1}^{\infty} (-1)^{n-1} n! \, \left ( \frac{4M^2}{Q^2}
\right )^n 
=  \frac{4M^2}{Q^2}  
 \left [ 1-  2 \,\left ( \frac{4M^2}{Q^2}  \right )
+6 \, \left (\frac{4 M^2}{Q^2} \right )^2 
+ {\cal O} (M^6/Q^6) \right ]
\ ,
\label{FG}
\end{equation}
   a result that has the  ${\cal O} (M^2/Q^2)$ 
   large-$Q^2$  behavior and structure similar to that of Eq. (\ref{FM}). 
   The crucial role of integration 
   over the $x \sim 1$  region in getting the power-law 
   behaviour is evident: 
   if the integration is restricted 
   to $x <x_0$, the outcome  vanishes exponentially (like 
   $\exp [-\bar x_0 Q^2 /4 x_0 M^2]$)   for large $Q^2$. 
   Thus, the   GPD ${\cal F}_G (x,Q^2)$ 
 corresponding to the Gaussian wave function 
 has 
 the same ${\cal F} (x,Q^2) = f(x) \, e^{-Q^2 g(x)}$
 structure as those considered in Refs. 
 \cite{Belitsky:2003nz,Diehl:2004cx,Guidal:2004nd}.

  \section{ Meromorphization } 
  
 One may question the  conjectures of  
 the model of Ref. \cite{Brodsky:2006uq}, e.g., the   interpretation 
 of the holographic variable $\zeta$ 
  as a  particular  product $\sqrt{x \bar x} b$ of  light-cone variables. 
We are going to show that the picture similar 
to that of Ref. \cite{Brodsky:2006uq} emerges also within the 
approach  related 
to  Migdal's program \cite {Migdal:1977nu}  
of Pad\'e  approximating the correlators $\Pi (p^2)$ of hadronic currents 
calculated in  perturbation theory.  
    Recently \cite{Erlich:2006hq}, it was demonstrated that some of Migdal's  results 
 coincide with those  of the holographic approach. 
Migdal's program  involves ``meromorphization''

\begin{equation} 
\Pi (p^2) = \frac1{\pi} \int_0^{\infty} \frac{\rho(s) }{s-p^2} \, ds \label{DisPi}
 \Rightarrow \Pi_{\cal M} (p^2) = \Pi (p^2) - \frac1{\pi {\cal Q} (p^2)} \int_0^{\infty} 
\frac{\rho (s) \, {\cal Q} (s)}{s-p^2} \, ds  \label{DisMig}
\end{equation} 
that  substitutes  the original
correlator $\Pi (p^2)$    by 
a   function   $ \Pi_{\cal M} (p^2)$  in which the  cut of the original correlator 
for real positive $p^2$ is eliminated by the second term in Eq. (\ref{DisMig}), with  
zeros of ${\cal Q} (p^2)$ at  timelike 
  $p^2$ generating the poles interpreted as  
  hadronic bound states.  Explicit Pad\'e  construction 
  in case of  correlators of currents   like  the scalar current 
  $\varphi   \varphi$
of scalar fields, vector current 
$\bar \psi \gamma_\mu \psi$ of spin-1/2 quarks, etc., gives
 $
{\cal Q} (p^2) \Rightarrow J_0 (\beta_{0,1} \sqrt{p^2}/M)  ,
$
 with 
$M$ being  the mass of the lowest state.
 In the deep spacelike region ($p^2 \equiv - P^2 \to - \infty $), 
  the difference between the 
  original expression and the approximant then 
  vanishes exponentially like $e^{-2\beta_{0,1} P/M}$.
   The coupling constant $f_M^2$ of the lowest state is given by
\begin{equation} 
 f_M^2 = \lim_{p^2 \to M^2}   (M^2 -p^2) \, \Pi_{\cal M} (p^2) = 
 \frac1{\pi {\cal Q}'(M^2)} 
 \int_0^{\infty} 
\frac{\rho (s) \, {\cal Q} (s)}{s-M^2} \,  ds  \label{fM2}  \ .
\end{equation} 
In the lowest order,  $\rho (s) = \rho_0 \, \theta (s)$, with $\rho_0 = 1/16 \pi$
 for $j=\varphi   \varphi$,  and 
$\rho_0 =N_c/12 \pi$ for vector and axial currents $\bar u \gamma_\mu (\gamma_5)d$
of spin-1/2 quarks. This gives  
\begin{equation} 
 f_M^2 =  
 \frac{  2 \rho_0 M^2}{\pi \beta_{0,1} J_1 (\beta_{0,1}) } 
 \int_0^{\infty} 
\frac{J_0 (\beta_{0,1} \sqrt{s}/M)}{M^2 -s} \,  ds 
=
 \frac{ 4 \rho_0 M^2}{\pi [\beta_{0,1} J_1 (\beta_{0,1}) ]^2} 
 \label{fM2f}  \ .
\end{equation}

For     form factors, it was suggested \cite{Dosch:1977qh} 
to     ``meromorphize''  the 
3-point function ${ T} (p_1^2, p_2^2,Q^2)$.
The lowest-order  triangle diagram has only the 
double spectral density $\rho(s_1,s_2,Q^2)$. 
 Building the function 
\begin{equation} 
{\cal T} (p_1^2, p_2^2,Q^2) = { T} (p_1^2, p_2^2,Q^2) + 
\frac1{\pi^2 {\cal Q} (p_1^2) {\cal Q} (p_2^2)} \int_0^{\infty} ds_1 \int_0^{\infty} ds_2 \ 
\frac{ \rho(s_1,s_2,Q^2)\, {\cal Q} (s_1) \, {\cal Q} (s_2) }{(s_1-p_1^2)(s_2-p_2^2)} \ ,
 \label{Trim}
\end{equation} 
 one removes the cuts in the $p_1^2$ and $p_2^2$ channels
 substituting them by poles  at the same locations 
 as in  $\Pi_{\cal M} (p^2)$. 
From this expression, one can extract the elastic form factor 
of the lowest state by using
\begin{eqnarray} 
 f_M^2 F_M(Q^2)
  &=& 
 \lim_{p_1^2 \to M^2}  \lim_{p_2^2 \to M^2} 
   (p_1^2 -M^2) (p_2^2 - M^2)\, 
 {\cal T} (p_1^2, p_2^2,Q^2) 
 \nonumber \\   &=&
 \frac1{\pi^2 [{\cal Q}'(M^2)]^2 } 
 \int_0^{\infty} ds_1 \int_0^{\infty} ds_2 \ 
\frac{ \rho(s_1,s_2,Q^2)\, {\cal Q} (s_1) \, {\cal Q} (s_2) }{(s_1-M^2)(s_2-M^2)} 
  \label{fM2F0}  \ .
\end{eqnarray}

 The  spectral densities $\rho (s_1,s_2,Q^2)$ 
can be calculated \cite{Radyushkin:1995pj,Radyushkin:2004mt} 
using the Cutkosky rules and light-cone variables in the frame where
the initial momentum $p_1$ has no transverse
components $p_1= \{p_1^+ = {\cal P}, p_1^- = s_1/{\cal P},{\bf 0}_{\perp}\} $,
while the momentum transfer $q \equiv p_2-p_1$ has  no
``plus'' component, 
$p_2= \{{\cal P}, (s_2+{\bf q}_{\perp}^2)/{\cal P}, {\bf q}_{\perp}\} $: 
\begin{equation}
\rho(s_1,s_2,Q^2) =\rho_0 \int_0^1 \,  dx \,  \frac{n(x)}{x\bar x}
 \,
\int d^2 {\bf k}_{\perp}
\,  \delta \left (s_1 - {{{\bf k}_{\perp}^2}\over{x \bar x}} \right )
\delta \left (s_2 - {{(k_{\perp}+ \bar x{\bf q}_\perp )^2}\over{x \bar x}} \right ) \  ,
\label{rhoF}
\end{equation}
where  $x$ is the fraction of ${\cal P}$ carried by the 
 active  quark,  and  ${\bf k}_{\perp}$ is 
 its transverse momentum; $\rho_0$ is  the same 
as in Eq. (\ref{fM2f}). 
For the simplest correlator  of three  scalar $\varphi \varphi$ currents, 
the numerator  factor
is 
$n(x) =1/x$.  
Taking   $i\varphi \stackrel{\leftrightarrow}{\partial^\mu}  \varphi$ 
for the EM  vertex gives $n(x)=1$  since 
$i \stackrel{\leftrightarrow}{\partial^\mu }\to  x {\cal P}^\mu$. Then  
\begin{eqnarray} 
 f_M^2 F_M^{\rm scalar} (Q^2) &=&  
 \frac{\rho_0}{\pi^2 [{\cal Q}'(M^2)]^2 } 
 \int_0^1 \, \frac{dx}{x\bar x}  \,
\int d^2 {\bf k}_{\perp}
\,  
\frac{{\cal Q} ({{{\bf k}_{\perp}^2}/{x \bar x}})}
{ M^2- {{{\bf k}_{\perp}^2}/{x \bar x}}} 
\  \frac {{\cal Q} ({{({\bf k}_{\perp}+ \bar x{\bf q}_\perp )^2}/{x \bar x}}) }
{M^2-{{({\bf k}_{\perp}+ \bar x{\bf q}_\perp )^2}/{x \bar x}}}  
  \label{fM2F}  \ ,
\end{eqnarray} 
i.e.,  a LC form factor expression.
   Using ${\cal Q} (s) = J_0 (\beta_{0,1} \sqrt{s}/M)$ and Eq. 
   (\ref{fM2f}) for $f_M^2$ , we obtain that the relevant wave  function
$\widetilde \Psi_{\cal M}^{\rm scalar} (x, {\bf k}_\perp)$  coincides with 
the momentum version (\ref{kperp})
of the holographic  wave function 
of Ref.\cite{Brodsky:2006uq},  supporting    
interpretation  in terms of light-cone variables $x,b$ proposed  there.

\section{ Spinor case}

\begin{figure}[h]
\mbox{\epsfysize=4cm
  \hspace{0cm}
  \epsffile{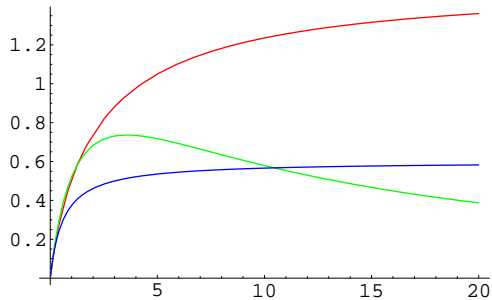}}
\caption{ Meson form factors: $Q^2 F_M (Q^2)$ (upper curve, red online),
$Q^2 F_M^{\rm spinor} (Q^2)$  (middle curve, green online); for comparison also shown 
$Q^2/(1+Q^2/m_\rho^2)$  (lower curve, blue online). 
 }
\label{mesonFF}
\end{figure}

Using  spin-1/2  quarks and  vector currents 
$j_\alpha, j_\beta$ for hadronized vertices,
one should deal with the amplitude $T^{\mu}_{\alpha \beta} (p_1,p_2)$ 
and  choose which tensor structure to consider. 
As a simple example, let us take  the projection 
$T^{\mu}_{\alpha \beta} n_\mu n^{\alpha} n^{\beta}$,  where 
$n$ is a lightlike vector having only the minus component 
in the frame specified above. For $\rho$-meson form factors, this projection picks out 
the combination  $C(Q^2) \equiv F_1 (Q^2) +\kappa F_2 (Q^2)   -\kappa^2 F_3  (Q^2) $, 
where $\kappa \equiv Q^2/2 m_\rho^2$.
For comparison, pQCD expects that the leading $1/Q^2$ behavior 
is provided by $F_{LL} (Q^2)  = F_1 (Q^2)  -\kappa F_2 (Q^2) 
 +(\kappa^2 +2 \kappa)  F_3  (Q^2) $,
with $F_1 (Q^2) \sim F_2 (Q^2) \sim 1/Q^4$ and $F_3 (Q^2) \sim 1/Q^6$ (see, e.g.,
\cite{Ioffe:1982qb}). 
Thus, $-C (Q^2)$  differs from    $F_{LL} (Q^2)$ only by  
$F_1 (Q^2)$ and $\kappa  F_3  (Q^2)$
terms which are considered in pQCD  as nonleading. 
The simplicity of this projection is that, for  a spinor triangle, 
the   numerator trace is given by the product of  quark 
light-cone ``plus'' momenta $x {\cal P},x {\cal P} $ and 
 $\bar x {\cal P} $. 
As a result, 
 $ n(x) = 6 x \bar x$. 
As we argued above, an  extra $\bar x$  factor in 
GPD  
${\cal F} (x, Q^2)$ should result in the  
   $1/Q^4$ behavior 
of the form factor at large $Q^2$.  To check this, we 
switch back  to the impact parameter space representation and 
observe  that 
 the  integral 
given by  Eq. (\ref{J0int}) changes  into 
\begin{equation} 
6 \int_0^1 dx \, x  \bar x \, J_0\left( \sqrt{\frac{1-x}{x}} \,  z \, Q \right) = 
\frac32 \,{z^2 Q^2}\, K_2(z Q) - \frac14 \, {z^3 Q^3} K_3(z Q) 
\equiv  {\cal K}_2 (zQ) \ . 
\label{J0int2}
\end{equation}
 Hence, the form factor is   
 given by an expression similar 
to Eq. (\ref{F0AdS}), but with 
${\cal K}_1(\xi)$ substituted by 
${\cal K}_2(\xi)$. Both functions are 
equal to 1  for $\xi=0$  
and behave as  $e^{-\xi}$
at   large $\xi$.  Thus,  
 one may  expect that $F_M^{\rm spinor} (Q^2)$ also 
  behaves like $1/Q^2$  for large $Q^2$.
 But  explicit calculation gives
\begin{eqnarray} 
F_M^{\rm spinor} (Q^2)
&=&  \frac{2 M^2}{Q^2  [\beta_{0,1}J_1(\beta_{0,1})]^2} 
\int_0^{\infty} {d\xi} \,  \xi \, 
 {\cal K}_2 (\xi )   \left [ 1- \frac12 \, {\xi^2}
\frac{M^2}{Q^2} 
 +\frac{3}{32}\,\xi^4 \frac{M^4}{Q^4}  - \frac{5}{576}\,\xi^6 \frac{M^6}{Q^6}
    + {\cal O} (M^8/Q^8) \right ] + {\cal O} (e^{-Q/\Lambda}) 
    \nonumber \\ &=& 
\frac{2M^2}{Q^2 [\beta_{0,1}J_1(\beta_{0,1})]^2}
 \left [ 0+ 24\, \frac{M^2}{Q^2} 
 - 288\, \frac{M^4}{Q^4}  + 2400\, \frac{M^6}{Q^6} 
    + {\cal O} (M^8/Q^8) \right ]+ {\cal O} (e^{-Q/\Lambda})
 \ . 
\label{F0MM2}
\end{eqnarray}
 The first term  vanishes because
$
 \int_0^{\infty} {d\xi} \, \xi \, 
{\cal  K}_2 (\xi) =0
$, 
and the leading term has  $1/Q^4$ behavior.
 However, the next, $1/Q^6$,   correction 
exceeds it 
for all $Q^2$ up to  $12M^2 \sim 7\,$\, GeV$^2$,
and  the $1/Q^4$ asymptotics establishes 
somewhere above 20\,GeV$^2$.   
Now,   this fact is very welcome: due to it, 
  the $F_M^{\rm spinor} (Q^2)$ 
curve in the region of a few GeV$^2$  imitates
the ``power counting'' $1/Q^2$ behavior much more 
successfully  than  Eq. (\ref{F0AdS}) that 
displays its nominal $1/Q^2$  asymptotics only well outside the 
few GeV$^2$ region (see Fig. \ref{mesonFF}).

 The  miraculous cancellation of the  moments for $K_2(\xi)$ and $K_3(\xi)$
in (\ref{F0MM2})  producing  the 
$ F_M^{\rm spinor} (Q^2)  \sim M^4/Q^4$ result
can be traced to 
 the $1/Q^4$ asymptotic behavior of the underlying 
double spectral density   $\rho (s_1,s_2, Q^2)$. 
To this end,  consider 
the double Borel transform of the three-point function 
 \cite{Ioffe:1982qb,Nesterenko:1982gc}

\begin{equation}
\Phi(\tau_1,\tau_2,Q^2) =\frac1{\pi^2}\int_0^\infty ds_1\int_0^\infty ds_2
 \ \rho(s_1,s_2,Q^2) \, e^{-s_1 \tau_1 -s_2 \tau_2} \ , 
\label{eq:doublebor}
\end{equation}
in  which the power weights 
are substituted by the exponential 
ones. 
 For  the triangle diagram
  \cite{Nesterenko:1982gc}, 
 \begin{equation} 
 \Phi(\tau_1,\tau_2,Q^2) = 
  \frac{N_c }{ 2\pi^2 (\tau_1 + \tau_2)} 
\int_0^1 dx \,  x \bar x \, 
\exp \left [ - Q^2\frac{\bar x \, \tau_1 \tau_2}{x(\tau_1+\tau_2)} \right ]  \  . 
 \label{Phipert2}
\end{equation}
 It contains  the $x\bar x$ factor (absent in the case of scalar
 quarks) that results in the $1/Q^4$ behavior of 
 $\Phi(\tau_1,\tau_2,Q^2)$ and, hence, 
 of the spectral density $\rho(s_1,s_2,Q^2)
 = {N_c} \,\theta(s_1)\, \theta (s_2) \,  (s_1+s_2)/2 Q^4 +\ldots$ \ .

\section{ Local quark-hadron duality}

Referring to 
the ``established'' $1/Q^2$ behavior of meson form factors at large $Q^2$,
one primarily has in mind the data on  the pion EM  form factor 
which resemble  the  VMD  monopole 
form $F_\pi (Q^2) \sim 1/(1+Q^2/m_\rho ^2)$.  
In fact, the data are  well  described 
by our local quark-hadron duality model  \cite{Nesterenko:1982gc} that 
incorporates, in a simplified form,  some ideas of  Migdal's 
program and the QCD sum rule approach \cite{Shifman:1978bx}.  
From the latter, we borrow the observation  that 
 only the lowest state  is narrow, and  use  the model  spectrum: 
   first 
 resonance 
 plus perturbative ``continuum''  starting from some effective 
 scale $s=s_0$.   
  In other words, we transform
   the   two-current correlator $\Pi (p^2)$   into

 \begin{equation} 
\Pi (p^2) = \frac1{\pi} \int_0^{\infty} ds \, \frac{\rho^{\rm pert} (s) }{s-p^2} 
 \Rightarrow \Pi^{\rm LD}  (p^2) = \frac{F_M^2}{M^2-p^2}  + \frac1{\pi}  \int_{s_0}^{\infty} 
\frac{\rho^{\rm pert}  (s)}{s-p^2} \, ds  \label{DisSR}  \,  . 
\end{equation} 
Then  we try   to reach the best possible agreement 
 between the original and model correlators  in the deep spacelike 
 region of $p^2 \equiv -P^2$. 
 For  axial currents, 
 $F_M \to f_\pi$, $M\to m_{\pi}\approx 0$, $\rho^{\rm pert} (s)  =1/4\pi$,
  and 
  \begin{equation} 
 \Pi (p^2) -\Pi^{\rm LD}  (p^2) = \frac{f_\pi^2}{p^2}   +  \frac1{4\pi^2}  \int_0^{s_0} ds \, 
\frac{\rho^{\rm pert} (s) }{s-p^2}  .  \label{DiffSR}
\end{equation} 
To  eliminate 
the leading $1/p^2$ term in this difference,  we should take 
 $ f_\pi^2 = s_0 /4 \pi^2$. As a result, the  coupling  $f_\pi$ 
  and the value of $s_0$ 
  are connected by  the {\it local  duality}
 relation  
 \begin{equation}
 \int_0^{s_0} \rho_\pi (s) \, ds = \int_0^{s_0} \rho^{\rm pert}(s) \, ds 
 \end{equation} 
  between the spectral density $\rho_\pi (s) = \pi f_\pi^2 \delta (s)$ 
 of the lowest state 
 and the perturbative  spectral density $\rho^{\rm pert}(s)$, applied  
 within the ``duality interval'' $s_0$. 
 Transforming the $n^\alpha n^\beta n_\mu$ projection of the 
 3-point function
 \begin{equation} 
{ T}^{\rm pert} (p_1^2, p_2^2,Q^2)  \Rightarrow 
\frac{f_\pi^2 \, F_\pi (Q^2)}{p_1^2 \, p_2^2}
 +  \frac1{\pi^2} 
  \int_0^{\infty} ds_1 \int_0^{\infty} ds_2 \  \Biggl [ 1- \theta (s_1 \leq s_0)
  \theta (s_2 \leq s_0) \Biggr ] \, 
\frac{ \rho^{\rm pert}(s_1,s_2,Q^2) }
{(s_1-p_1^2)(s_2-p_2^2)} \  ,
 \label{TriSR}
\end{equation} 
 and requiring that the difference between $T^{\rm pert} (p_1^2, p_2^2,Q^2)$
 and the model function vanishes faster than $1/p_1^2 p_2^2$ for large spacelike
 $p_1^2, p_2^2$, gives the 
local duality relation for the pion form factor 
\cite{Nesterenko:1982gc}
\begin{equation}
f_{\pi}^2F_{\pi}^{\rm LD}(Q^2) = \frac1{\pi^2}\int_0^{s_0}ds_1
\int_0^{s_0} ds_2 \ \rho^{\rm pert}(s_1,s_2,Q^2).
\label{FpiLD}
\end{equation}
Using Eq.   (\ref{rhoF}) for  $\rho^{\rm pert}(s_1,s_2,Q^2)$ with 
$n(x) = 6x \bar x$
and $f_\pi^2 = s_0 /4 \pi^2$ 
produces   the  Drell-Yan  
formula 
\begin{equation}
F_{\pi}^{\rm LD}(Q^2) =  \frac{6 }{\pi s_0}
 \int_0^1 \,  dx \,  
\int d^2 {\bf k}_{\perp}
\,  \theta \left ({\bf k}_{\perp}^2 \leq  x \bar x  s_0 \right )
\theta  \left ( ({\bf k}_{\perp}+ \bar x{\bf q}_\perp )^2  \leq x \bar x s_0
 \right ) 
\label{DYW}
\end{equation}
with the effective 
``local duality  wave function''  
$\widetilde \Psi^{\rm LD}_\pi (x,{\bf k}_{\perp}) = \sqrt{{6 }/{\pi s_0} } \
\theta({ k}_{\perp}^2\leq  x\bar x s_0) $ for the pion. 
In the  impact parameter representation,  
$\Psi^{\rm LD}_\pi  (x, {\bf b}_\perp ) = 
\sqrt{{6 x \bar x}/{\pi}}
 \,
J_1(b \sqrt{x \bar x  s_0 })/b $.
Taking  the  $b \to 0$ limit   gives the model distribution amplitude   
 $\varphi^{\rm LD}_\pi (x)= 6 f_\pi x\bar x$, that 
coincides with the asymptotic pion DA. 
 Explicit expression for the pion form factor in the local duality model is known from 
\cite{Nesterenko:1982gc}:
\begin{equation} 
 F_{\pi}^{\rm LD} (Q^2) =  1- \frac{1+{6s_0}/{Q^2}}
 {\left (1+{4s_0}/{Q^2} \right )^{3/2}}
 = \frac{6s_0^2}{Q^4} - \frac{40s_0^3}{Q^6} + \frac{210s_0^4}{Q^8} - \frac{1008s_0^4}{Q^8}
+  \  {\cal O} (s_0^5/Q^{10})  \ . \label{FLDor}
\end{equation}
As expected, $F_{\pi}^{\rm LD} (Q^2)$ behaves like  $1/Q^4$ for large $Q^2$, 
but the expansion parameter of this series 
is again large $\sim 6s_0/Q^2 \approx 4$\,GeV$^2$, so that 
the asymptotic $1/Q^4$ estimate  should not be used at   accessible $Q^2$.   
However, in the region of a few $Q^2$, the $F_{\pi}^{\rm LD} (Q^2)$ curve  successfully imitates
the  $1/Q^2$ behavior and goes very close to existing \cite{Volmer:2000ek}
and preliminary  \cite{Horn:2006} 
experimental data. 

\begin{figure}[h]
   \mbox{\epsfysize=4cm
  \hspace{0cm}
  \epsffile{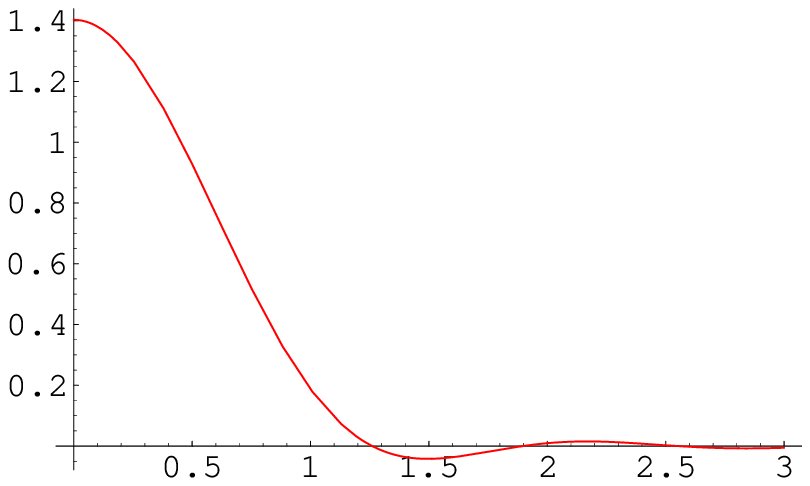} \hspace{1cm} \epsfysize=4cm
  \epsffile{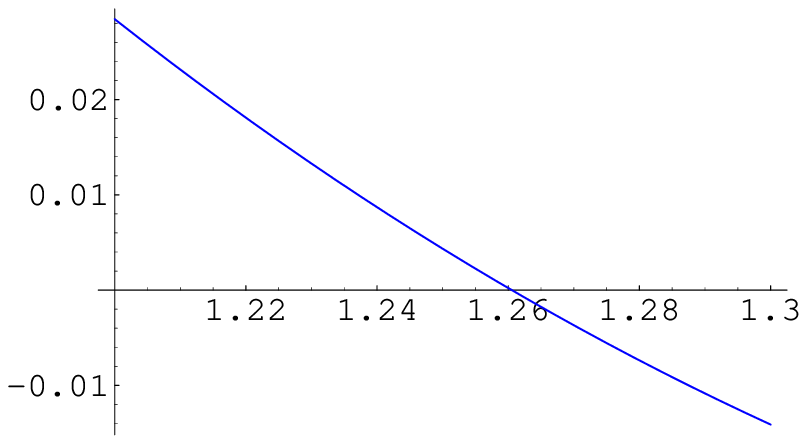}}
\caption{ ``Holographized'' local duality wave function
$\sqrt{\pi s_0/6} \, \Psi^{\rm LD,holo}_\pi (x,{\bf k}_{\perp})$ vs. $ k_{\perp}/\sqrt{x \bar x}$
in GeV.
 }
\label{holoLD}
\end{figure}

The  sharp cut-off form of 
$\Psi^{\rm LD}_\pi (x,{\bf k}_{\perp})$ is a consequence
of a simple model for the higher states' contribution.
The resulting $b$-space function  $\Psi^{\rm LD} (x, {\bf b}_\perp )$
has a Bessel-type form, but goes beyond the first zero.   
 ``Holographizing'' it by imposing the cut-off 
$\theta (b \sqrt{x \bar x  s_0 } \leq \beta_{1,1})$, we found  that  
its  ${\bf k}_{\perp}$ version  has a 
smooth dependence on ${ k}_{\perp}^2/x\bar x$,
qualitatively similar to that of Eq.(\ref{kperp}), with the lowest zero 
located at ${ k}_{\perp}^2/x\bar x \approx (1.26$\,GeV$)^2$, 
i.e., ``unexpectedly close'' to 
the $A_1$ position (see Fig. \ref{holoLD}).

\section{ Higher orders and transition to pQCD}

\begin{figure}[h]
\mbox{
   \epsfxsize=9cm
 \epsfysize=4cm
 \hspace{0cm}  
  \epsffile{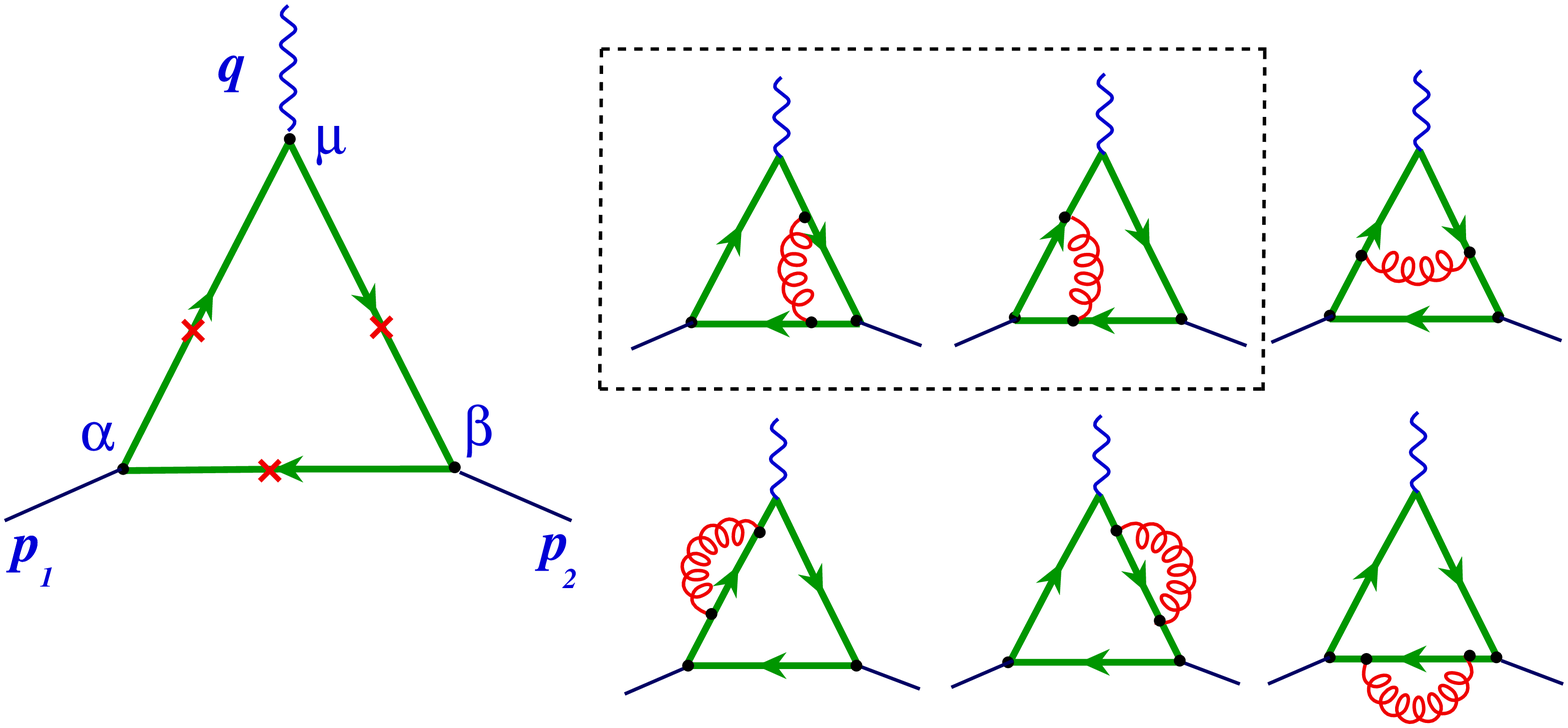}  } 
  \caption{ Diagrams producing spectral density $\rho  (s_1, s_2, Q^2)$
  at  the  one-loop and two-loop (${\cal O} (\alpha_s)$) level.
  Outlined diagrams are responsible for the leading 
  pQCD $1/Q^2$ behavior.}
\label{Figrho}
\end{figure}

Both in the meromorphization 
and  local duality 
approaches one can  include higher order $\alpha_s$ corrections 
to   spectral densities. 
  In particular, 
the two-loop calculation of $ \rho (s_1, s_2,Q^2)$ 
for axial and vector currents 
was performed in \cite{Braguta:2004ck}. 
Among ${\cal O}(\alpha_s)$  contributions, 
there are   gluon-exchange diagrams
whose asymptotic large-$Q^2$  behavior is determined 
by the hard pQCD mechanism 
(see Fig. \ref{Figrho}).
 As a result,  the leading $1/Q^2$ term 
of the spectral density 
 corresponding to the $T^{\mu}_{\alpha \beta} n_\mu n^{\alpha}
 n^{\beta}$ projection can be written in pQCD-like form,
 \begin{equation} 
\rho_{\alpha_s}  (s_1, s_2, Q^2) =
 2 {\pi}{\alpha_s}   \,  
\frac{C_F }{N_c} 
 \int_0^1 {dx} \,  \int_0^1 dy  \ 
\frac{\rho (x,s_1) \rho  (y,s_2)}{xyQ^2} 
 \,  + {\cal O} (1/Q^4) \, , 
 \end{equation}
where  $\rho (x,s_1)$ 
is the lowest order term for the  $x$-unintegrated spectral density,

 \begin{equation}
\rho (x,s_1) = \frac{N_c}{2\pi^2}  \int
 \delta \left (s_1- \frac{k_{1 \perp}^2}{x \bar x} \right ) 
 \, d^2 {\bf k}_{1 \perp}  =  \frac{N_c}{2\pi} \, \theta (s_1) \, 
 x \bar x  
\end{equation}

(and similarly for $\rho (y,s_2)$).

\begin{figure}[h]
\mbox{\epsfysize=4cm
  \hspace{0cm}
  \epsffile{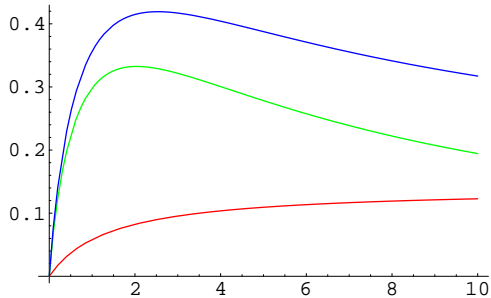}}
 \caption{ Pion form factor  in local quark-hadron duality model:
 $Q^2 F_\pi^{{\rm LD} (\alpha_s)}$  (lower curve, red online), 
$Q^2 F_\pi^{{\rm LD}}$  (middle curve, green online), 
total contribution (upper curve, blue online).
 }
\label{pionLD}
\end{figure}

 Integrating  $\rho (x,s)$ over the 
duality interval $0 \leq s \leq s_0$ and dividing result
by $\pi f_\pi$ gives  the local duality model 
$\varphi^{\rm LD}_\pi (x)= 6 f_\pi x\bar x = \varphi^{\rm as}_\pi (x)$
for the pion DA. 
 Thus, substituting the ${\cal O} (1/Q^2)$ part of 
$\rho_{\alpha_s}  (s_1, s_2, Q^2)$  
 into the local duality
 relation (\ref{FpiLD}) gives 
 the pQCD hard gluon exchange contribution
 $F_{\pi}^{\rm pQCD} (Q^2) = 8 \pi \alpha_s f_\pi^2 /Q^2$ 
 calculated for  the asymptotic shape of the pion DA. 
 It is instructive to rewrite this result  in the form 
 $F_{\pi}^{\rm pQCD} (Q^2) = 2 (s_0/Q^2)  (\alpha_s/\pi)$ that clearly 
 reveals 
 its nature as the $\alpha_s$  correction to the 
 soft contribution (\ref{FLDor}),
 with the $(\alpha_s/\pi)$  factor 
 being the standard penalty for an extra loop. 
 Using full expressions for  $\rho_{\alpha_s}  (s_1, s_2, Q^2)$ 
 one can get the local duality model prediction 
 for $F_\pi^{{\rm LD} (\alpha_s)}$ at all $Q^2$,
 see Ref. \cite{Braguta:2004ck}.  
 In fact, a very good approximation is given
 by a simple 
 interpolation
 formula $F_{\pi}^{{\rm LD} (\alpha_s)} (Q^2) =  
 (\alpha_s/\pi)/(1 +Q^2/2 s_0) $ \cite{Radyushkin:1990te}
 between the $Q^2=0$ value  
 $F_{\pi}^{{\rm LD} (\alpha_s)} (0 ) = \alpha_s/\pi$ 
 (that can be obtained, e.g., from the two-point 
 result $\rho (s) = \theta(s) (1+\alpha_s/\pi)$ 
 using the Ward identity) and the large-$Q^2$ 
 asymptotic behavior.  
 With $\alpha_s/\pi  \approx 0.1$, the ${\cal O}(\alpha_s)$ 
 term is a $\lesssim 30\%
 $  correction to 
 the ${\cal O}(\alpha_s^0)$ 
 term in the $Q^2 \leq 4$\, GeV$^2$ region  (see Fig. \ref{pionLD}), 
 and their sum is in  good  agreement with existing data.

 \section {Summary and conclusions} 
 
 In this letter, we 
 studied the large-$Q^2$ behavior of the 
 meson form factor $F_M (Q^2)$   constructed 
 using the holographic model 
 of Ref. \cite{Brodsky:2006uq}, and observed that,
 despite its $1/Q^2$ asymptotic behavior, 
 the combination $Q^2 F_M (Q^2)$  
 is  not  flat 
 in the  accessible region $Q^2 \lesssim 10$\, GeV$^2$.
 We also found that the asymptotic  $1/Q^2$ result
 is governed by the Feynman mechanism rather than by 
 large transverse momentum 
 dynamics.  
 We discussed  the meromorphization approach,
 in which the form factors are given 
 by integrating the perturbative spectral density 
  $\rho (s_1,s_2,Q^2)$  with weights proportional to  
  $J_0  (\beta_{0,1} \sqrt{s_i}/M) /(s_i-M^2)$, and   showed that, 
if one uses scalar currents made of scalar quarks,
 the result coincides with the expression 
 given in 
 Ref. \cite{Brodsky:2006uq}. 
  For  spin-1/2 quarks, we studied a particular tensor   
 projection 
 of the 3-point function, and demonstrated that,
 due to an extra $(1-x)$ factor,    
 the resulting form factor $F^{\rm spinor}_M (Q^2)$ 
 has $1/Q^4$ asymptotic 
 behavior. However, owing to the  late onset of this 
  asymptotic  pattern, the combination  
  $Q^2 F^{\rm spinor}_M (Q^2)$ is rather flat
  in the phenomenologically important few GeV$^2$
 region, i.e., $F^{\rm spinor}_M (Q^2)$
 imitates the $1/Q^2$ 
  behavior.  
  Then we presented the results for the 
 pion form  factor obtained  in the local quark-hadron 
 duality model, which  corresponds to 
  integrating the  perturbative spectral density 
  $\rho (s_1,s_2,Q^2)$  with $\theta (s_1 \leq s_0)$
  weights. Again, the lowest-order term 
  has nominally the $1/Q^4$ asymptotics,
  but it  imitates the $1/Q^2$ 
  behavior in the few GeV$^2$ region. 
  Including   the ${\cal O}(\alpha_s)$ correction 
  term of $\rho (s_1,s_2,Q^2)$ 
  brings in  the hard pQCD  contribution
  having the dimensional counting $1/Q^2$ behavior
  at large $Q^2$.  However, at  accessible $Q^2$, 
  the   ${\cal O}(\alpha_s)$ term  is a rather 
  small fraction of the total result, 
  because of small $\alpha_s/\pi \sim 0.1$  factor 
  associated with each higher order correction.  
  
  We did not discuss  the nucleons form factors  in the present paper,
  but we want to mention that  
   the lowest-order perturbative density 
  for spin-1/2 quarks is known \cite{Nesterenko:1982gc},
  and since the double Borel representation (see Eq.(\ref{Phipert2})) 
  in that case has $(1-x)^2$  factor,  the resulting  
   asymptotic behavior of  $\rho (s_1,s_2,Q^2)$ is $1/Q^6$.
   But the local duality result for $G_M^p(Q^2)$
   closely follows the dipole shape of the  data up to $Q^2 \sim 15$\, GeV$^2$. 
     
  Thus, we  observe that the power of $(1-x)$ in these perturbative 
  versions of the relevant parton densities $f(x)$  increases  with the number of quarks
  like $f_n(x) \sim (1-x)^{n-1}$, i.e.,  the probability that the total 
  momentum of $n-1$ spectators is $x_{\rm sp}$ goes like $x_{\rm sp}^{n-1}$,
  which looks quite natural. 
 Due to  Feynman mechanism,
  this formally gives $(1/ Q^2)^{n}$ behavior for form factors. 
  However, because of  the late onset of the asymptotic regime,
  the form factors imitate  $(1/Q^2)^{n-1}$ behavior 
  in  a rather wide preasymptotic region.  
 In this scenario, the ``observed'' quark counting rules is an approximate 
 and transitional phenomenon dominated by nonperturbative,
 long-distance aspects of hadron dynamics.

 {\it Acknowledgements.} I thank  S. J. Brodsky and G. 
 de Teramond for correspondence,
 and H. Grigoryan for useful  discussions. 
 
{\it Notice:} This manuscript has been authored by Jefferson
 Science Associates, LLC under Contract No.
 DE-AC05-06OR23177 with the U.S. Department of Energy.
 The United States Government retains and the publisher,
 by accepting the article for publication, acknowledges
 that the United States Government retains a non-exclusive,
 paid-up, irrevocable, world-wide license to publish or
 reproduce the published form of this manuscript, or allow
 others to do so, for United States Government purposes.

\end{document}